\documentstyle[12pt]{article}
\begin{document}

\renewcommand{\baselinestretch}{1.5}
\newcommand\beq{\begin{equation}}
\newcommand\eeq{\end{equation}}
\newcommand\bea{\begin{eqnarray}}
\newcommand\eea{\end{eqnarray}}

\def\com{centre of mass }
\def\es{exclusion statistics }
\def\gs{ground state }
\def\ms{mutual statistics }

\centerline{\bf A MULTISPECIES CALOGERO-SUTHERLAND MODEL}
\vskip 2 true cm

\centerline{Diptiman Sen \footnote{E-mail: ~diptiman@cts.iisc.ernet.in}} 
\centerline{\it Centre for Theoretical Studies, Indian Institute of 
Science,}
\centerline{\it Bangalore 560012, India}
\vskip 2 true cm

\noindent {\bf Abstract}
\vskip 1 true cm

Motivated by the concept of ideal mutual statistics, we study a multispecies 
Calogero-Sutherland model in which the interaction parameters and masses 
satisfy some specific relations. The ground state is exactly solvable if those 
relations hold, both on a circle and on a line with a simple harmonic 
potential. In the latter case, the one-particle densities can be obtained 
using a generalization of the Thomas-Fermi method. We calculate the second 
virial coefficients in the high temperature expansion for the pressure. We 
show that the low-energy excitations are the same as those of a Gaussian 
conformal field theory. Finally, we discuss similar relations between the 
statistics parameters and charges for a multispecies anyon model in a magnetic 
field.

\vskip 1 true cm
\noindent PACS Nos.: ~5.30.-d, ~71.45.Jp, ~74.20.Kk.
\vskip 1 true cm


\newpage

\leftline{\bf 1. Introduction}
\vskip .3 true cm

The Calogero-Sutherland (CS) model \cite{CAL,SUT} consists of a system of
identical particles in one dimension which interact with each other via 
a two-body inverse square potential. The model has been 
studied for many years from several points of view. It is exactly solvable and
integrable \cite{POL,MOS}, it has interesting connections with random matrix 
theory \cite{MEH}, and it provides an example of \es \cite{HAL1}. 

The concept of \es introduced by Haldane has attracted considerable 
attention recently [7-15]. The examples of {\it ideal} \es known so far include 
the Calogero-Sutherland (CS) model, spinons in the 
Haldane-Shastry model \cite{HAL2,POL}, and anyons residing in the lowest Landau 
level in a strong magnetic field [7-10]. The term {\it ideal} means that the 
only interactions between particles are statistical, and that these 
interactions are scale invariant, i.e., independent of the momentum or energy 
scale \cite{FUK}.

The idea of \es can be generalized from systems with a single species of 
indistinguishable particles to systems with several species. In the latter 
case, the concept of {\it mutual statistics} has proved to be very useful 
[10-12,14]. An example of \ms is a multianyon system in two dimensions 
\cite{ISA1,WU}. It seems interesting to ask whether there is a one-dimensional
model which can realize \ms. With this and other motivations, a multispecies 
CS model was examined in Ref. \cite{FUR}; however the model involved both 
two-body and three-body interactions. In this paper, we ask whether it may 
be possible to realize \ms in a CS model with only two-body interactions. 
Somewhat unexpectedly, we discover that ideal \ms cannot be realized
unless the interaction parameters and masses of the different species satisfy 
certain relations. However we are not able to prove that the model does
exhibit \ms if those relations {\it are} satisfied.

The outline of our paper is as follows. In Sec. 2, we present a general 
analysis of a model satisfying ideal \ms on a line. We assume that the \ms is 
described by a {\it symmetric} matrix. (This assumption is motivated by a CS 
model). We find that the masses and interaction parameters must satisfy certain 
relations in order that the asymptotic Bethe ansatz be applicable to the 
ground state. In Sec. 3, we consider a CS model which satisfies 
those relations. It is known that the \gs is exactly solvable for the model on 
a circle; it is also exactly solvable if the CS model is placed on a line 
in an external simple harmonic 
potential, with the same harmonic frequency for all species \cite{KRI}. We
discuss a class of exactly solvable excited states, including some
subtleties involving the \com excitations on a circle. In 
Sec. 4, we compute the one-particle densities in the harmonic potential 
problem using a generalization of the Thomas-Fermi method. In Sec. 5, we 
study the CS model at high temperature, and calculate the second virial 
coefficients in a high temperature expansion of the equation of state. In Sec. 
6, we study the low-energy spectrum and eigenstates to leading order in $1/L$. 
We find that the spectrum is the same as that of a $c=1$ Gaussian conformal 
field theory. We derive the leading terms in the specific heat and pressure
at low temperature. Finally, in Sec. 7, we consider a multispecies anyon 
model in an uniform magnetic field at zero temperature and show that it has 
some desirable properties if the {\it charges} and statistics parameters 
satisfy the same kinds of relations \cite{ISA1}.

\vskip .7 true cm
\leftline{\bf 2. Mutual Exclusion Statistics in One Dimension}
\vskip .3 true cm

Given a many-body quantum system with a finite number of states and $A$
species of particles, let $N_a$ denote the number of particles of species $a$ 
($a=1,2,...,A$). Let $D_a$ be the number of states which are available to the 
next $a$ particle which may be added to the system. The $D_a$'s are functions
of the numbers $N_a$. The \ms parameters $g_{ab}$ are defined as
\beq
\Delta D_a ~=~ -~ \sum_b ~g_{ab} ~\Delta N_b ~.
\label{1}
\eeq
If the diagonal matrix element $g_{aa} = 0$ (or $1$), the $a$ particles are 
bosons (or fermions). We will henceforth assume that all $g_{ab} \ge 0$. In 
the literature, the {\it off-diagonal} element $g_{ab}$ is referred to as the 
`mutual' statistics between species $a$ and $b$. 

For an ideal system with a single species such as a CS model, the parameter 
$g =g_{aa}$ can be interpreted in terms of the phase shift which appears 
in a two-particle scattering \cite{BER}. Namely, the scattering phase shift is 
given by $\delta (p)=\pi g \theta (p)$, where $p$ is the relative momentum of 
the two particles, and $\theta (p)=p/\vert p \vert$ for $p \ne 0$. (Thus the 
wave function of two bosons is symmetric with $g=0$, and that of two fermions 
is antisymmetric with $g=1$). At zero temperature, the particles occupy momenta
satisfying
\beq
p_i ~=~ {{\pi \hbar g} \over L} ~\sum_{j \ne i} ~\theta (p_i - p_j) ~,
\label{dist1}
\eeq
where $L$ is the length of the system. Eq. (\ref{dist1}) may be derived by an 
asymptotic Bethe ansatz analysis of the CS model on a circle of length $L$ 
\cite{SUT,POL,YAN}. Thus momenta from $-p_F$ to $p_F$ are occupied with a 
spacing of $2 \pi \hbar g/L$. If $N$ is the number of particles, we have 
\beq
\int_{-p_F}^{p_F} ~dp~ {L \over {2 \pi \hbar g}} ~=~ N ~.
\label{2}
\eeq
We are eventually interested in the thermodynamic limit $N, L \rightarrow 
\infty$ keeping the particle density $\rho =N/L$ fixed. Then 
\beq
p_F ~=~ \pi \hbar g \rho ~,
\label{3}
\eeq
and the total energy 
\beq
E ~=~ \int_{-p_F}^{p_F} ~dp~ {L \over {2 \pi \hbar g}} ~{{p^2} \over {2m}} ~=~ 
{{\pi^2 \hbar^2 g^2} \over {6m}} ~L \rho^3 ~.
\label{3p1}
\eeq

We may now generalize the phase shift argument to a multispecies model. Let us
assume that a scattering between two particles $i$ and $j$ produces a phase 
shift $\pi g_{ij} \theta (v_i - v_j)$, where $v=p/m$ is the velocity, $m_i 
m_j /(m_i + m_j)$ is the reduced mass, and $(v_i - v_j) m_i m_j /(m_i + m_j)$
is the relative momentum of the two particles. The two particles may or may 
not belong to the same species; $g_{ij} = g_{ab}$, $m_i = m_a$ and $m_j = m_b$ 
if $i$ and $j$ belong to species $a$ and $b$ respectively. (It is clear that 
$g_{ab}=g_{ba}$ in such a model). If we assume the asymptotic Bethe ansatz to 
be valid for the \gs of this model, we obtain 
\beq
m_i v_i ~=~ {\pi \hbar \over L} ~\sum_{j \ne i} ~g_{ij} ~\theta (v_i - v_j ) ~.
\label{dist2}
\eeq
We assume that no two particles have the same velocity; this will be verified
in Eq. (\ref{7p1}) below.

If $N=\sum_a N_a$, Eq. (\ref{dist2}) has $N !\ / \prod_a N_a !\ $ distinct 
solutions; this is because the particles $1,2,...,N$ can be ordered with 
increasing velocities in that many different ways. (Recall that different 
orderings of identical particles are physicaly indistinguishable). The
asymptotic Bethe ansatz \gs wave function will then be a superposition of that 
many different plane waves. For this idea to work, it is clearly essential that 
the total energy $E = \sum_i m_i v_i^2 /2$ be the same for all the 
different solutions of (\ref{dist2}). We find that this is true if and only if
\beq
{{g_{ij} g_{ik}} \over {m_i}} ~=~ {{g_{ij} g_{jk}} \over {m_j}}
\label{gm1}
\eeq
for any three different particles $i, j$ and $k$; they may or may not belong to
the same species. The non-trivial solution of (\ref{gm1}) is given by
\beq
g_{ij} ~=~ \alpha ~m_i m_j ~.
\label{gm2}
\eeq
Eq. (\ref{gm2}) is one of our main results. Note that the $A \times A$ matrix
$g_{ab}$ then has only one non-zero eigenvalue. 

In deriving (\ref{gm2}) from (\ref{gm1}), we have assumed that all the 
off-diagonal elements $g_{ab} > 0$. There {\it are} other solutions of 
(\ref{gm1}) in which an off-diagonal $g_{ab}$ vanishes. However in that 
case, (\ref{gm1}) implies that $g_{ac} g_{bc} =0$ for {\it all} 
$c$. We can then prove, by induction in the number $A$, that the system 
consists of two or more disjoint groups of species, such that $g_{ab} =0$
whenever $a$ belongs to one group and $b$ belongs to another. We can therefore
study each group separately. We will not consider such cases any further, and 
will assume henceforth that all the $g_{ab}$ satisfy (\ref{gm2}). 

It is rather surprising and novel that {\it mass-dependent} conditions such as 
(\ref{gm2}) appear to be necessary for mutual statistics. Such conditions may
be peculiar to models in which the two-particle scattering phase shift does not 
depend on the {\it magnitude} of $v_i - v_j$. We do not know of any physical 
system in which such relations might arise in a natural way. 

Using (\ref{gm2}), we can solve (\ref{dist2}) explicitly. If the 
particles are ordered so that $v_1 < v_2 < ... < v_N$, we get
\beq
v_i ~=~ {{\pi \hbar \alpha} \over L} ~\sum_{j \ne i} ~m_j ~ \theta (i-j) ~.
\label{7p1}
\eeq
From Eqs. (\ref{dist2}) and (\ref{gm2}), we deduce the \gs energy to be
\beq
E_o ~=~ {{\pi^2 \hbar^2 \alpha^2} \over {6 L^2}} ~\Bigl[ ~(~\sum_a ~m_a 
N_a ~)^3 ~-~ \sum_a ~m_a^3 N_a ~\Bigr] ~.
\label{3p5}
\eeq
The Fermi momentum for species $a$ may be obtained in two different ways. 
Firstly, for an ideal system at zero temperature, the chemical potential 
$\mu_a = p_{Fa}^2 / 2 m_a$. Further, $\partial E_o / \partial N_a =\mu_a$. In
the thermodynamic limit, the Fermi momentum for species $a$ is therefore given 
by
\beq
p_{Fa} ~=~ \pi \hbar ~\sum_b ~g_{ab} ~ \rho_b ~,
\label{4}
\eeq
where $\rho_b = N_b / L$. We observe that the Fermi velocity 
\beq
v_F ~=~ {{p_{Fa}} \over {m_a}} ~=~ \pi \hbar \alpha ~\sum_b ~m_b ~\rho_b 
\label{fv}
\eeq
is the same for all species. The second way of deriving $v_F$ is to examine 
(\ref{dist2}); amongst all the solutions of that equation, the one with the 
largest possible value of $v_a$ is clearly given by (\ref{fv}). We note that
the \gs pressure and compressibility are given by
\bea
P ~&=&~ - ~{{\partial E_o} \over {\partial L}} ~=~ {{v_F^3} \over {3 \pi \hbar
\alpha}} ~, \nonumber \\
\kappa ~&=&~ -~ \Bigl( L ~{{\partial P} \over {\partial L}} \Bigr)^{-1} ~=~
{{\pi \hbar \alpha} \over {v_F^3}} ~.
\label{press}
\eea

We can understand Eq. (\ref{1}) semiclassically by identifying the number of 
states with the `volume' of phase space. The volume of phase space occupied 
in the \gs by particles of species $a$ is given by $\int (dxdp/2\pi
\hbar) [1+\theta (\mu_a - h_a)]/2$, where $h_a =p^2/2m_a $ is the 
single-particle Hamiltonian. The volume available to the next $a$ particle is 
then given by the `remaining' amount
\beq
D_a (\epsilon) ~=~ \int {{dx dp} \over {2 \pi \hbar}} ~{1 \over 2}[ 1 - \theta 
(\mu_a -h_a )] ~\exp (-\epsilon h_a ) ~,
\label{4p2}
\eeq
where we have introduced a cutoff $\epsilon$ to make sense of the divergent
integral; we will let $\epsilon \rightarrow 0^+$ at the end
of the calculation. (The final result should not depend on the cutoff
procedure \cite{MUR1}). From Eq. (\ref{4}), the change in Fermi momentum is
given by
\beq
\Delta p_{Fa} ~=~ {{\pi \hbar} \over L} ~\sum_b ~g_{ab} ~\Delta N_b ~.
\label{4p3}
\eeq
Combining Eqs. (\ref{4p2}-\ref{4p3}), we find that $\Delta D_a \equiv 
\lim_{\epsilon \rightarrow 0} \Delta D_a (\epsilon)$ is given by Eq. 
(\ref{1}). 

In the entire discussion above, we have implicitly assumed the system to be 
homogeneous, that is, the densities $\rho_a$ do not depend on the position $x$. 
This is a non-trivial assumption for the \gs of a multispecies model. 
For instance, consider a two-species model in which $g_{12}$ is much larger 
than both $g_{11}$ and $g_{22}$. Then it is intuitively clear that the \gs 
will exhibit phase separation, i.e., the particles belonging to species 
$1$ and $2$ will prefer to reside in two different regions of apace. One would 
then use Eq. (\ref{3}) in the two regions separately, where $\rho_a$ is the 
{\it local} density of the species appropriate to a particular region; it 
would be incorrect to use the multispecies Eq. (\ref{4}) where $\rho_a$ is the 
{\it average} density defined over the entire system.

One can derive the {\it leading} term in Eq. (\ref{3p5}) in a different way 
using 
the result (\ref{3p1}) for the case of a single species. For a CS model, the 
particles have such a singular two-particle interaction that no two of them 
can cross each other. (This will be explained in more detail in the next 
section). The Hilbert space thus breaks up into a number of different sectors 
corresponding to different orderings of the particles on the line. It then 
seems reasonable to demand that a thermodynamic quantity like the \gs energy 
should only depend on the numbers $N_a$, and not on which sector we are 
working in. This is equivalent to requiring that the \gs
should not show phase separation. Let us therefore calculate the \gs energy in 
the sector where the different species are completely segregated. The $N_a$ 
particles of species $a$ occupy a length $L_a$, so that $\sum_a L_a = L$. For 
a single species, we know that the \gs energy $E_{oa} = \pi^2 \hbar^2 g_{aa}^2 
N_a^3 / (6 L_a^2 m_a)$. Now we minimize $E_o = \sum_a E_{oa}$ as a function of 
the $L_a$ keeping the sum of the $L_a$ fixed; this is equivalent to saying 
that the pressures exerted by the different species must be equal. (In the 
thermodynamic limit, the contributions to the energy coming from the regions 
separating two species can be neglected). This calculation leads to the 
expression in (\ref{3p5}). For later use, we note that the product 
$m_a N_a /L_a$ is the same for all $a$ in this sector. In fact, the mass
density $\sum_a m_a \rho_a$ must have the same value at all positions and in 
{\it all} sectors due to Eqs. (\ref{fv}-\ref{press}), because the pressure must 
be the same throughout the system. However this sum rule does not fix the
{\it individual} densities $\rho_a$ which depend on the sector (see Sec. 4).

Since particles cannot exchange their positions in the CS model, two species 
are indistinguishable if their masses are equal. (Actually, this is only true
for the model on a line, not on a circle. For instance, for two particles on
a circle, we will show below that the energy spectrum is different for
identical and non-identical particles). We will assume that all the masses are 
unequal; if some of them are equal, the CS model on a line reduces to one with 
a smaller number of species.

\vskip .7 true cm
\leftline{\bf 3. A Multispecies Calogero-Sutherland Model}
\vskip .3 true cm

A CS model satisfying the relations in Eq. (\ref{gm2}) has another attractive 
feature, namely, that its \gs can be solved exactly \cite{KRI}. We first 
present the Hamiltonian for a general CS model defined on a circle of 
perimeter $L$.
\beq
H ~=~ \sum_i ~{{p_i^2} \over {2m_i}} ~+~ {{\pi^2 \hbar^2} \over {L^2}} ~
\sum_{i<j} ~g_{ij} (g_{ij} -1)~ {{m_i + m_j} \over {2m_i m_j}} ~{1 \over 
{\sin^2 [~{\pi \over L} ~(x_i - x_j) ~]}} ~,
\label{ham1}
\eeq
where the sum over $i,j$ runs over all the $N=\sum_a N_a$ particles. Since the 
Hamiltonian does not distinguish between $g_{ij}$ and $1-g_{ij}$, one needs to 
specify the boundary condition on the allowed wave functions, namely, that 
$\Psi \sim \vert x_i - x_j \vert^{g_{ij}}$ when $x_i - x_j \rightarrow 0$. 
Since the wave functions vanish at coincident points, it is clear that there 
are a number of disjoint sectors which correspond to different orderings of 
distinguishable particles as we go anticlockwise around the circle \cite{CAL}. 
Under time evolution, a configuration of particles cannot change from one 
sector to another. To illustrate the idea of sectors, consider a system with
two particles of species $1$ and two of species $2$. Then there are two
sectors, namely, $(1122)$ and $(1212)$. 

We will work in a {\it bosonic} basis; if all the $N$ particles belong to the 
same species, we demand that 
\beq
\Psi (0, x_1 , x_2 , ..., x_{N-1}) ~=~ \Psi (x_1 , x_2 , ..., x_{N-1}, L) ~,
\label{bc1}
\eeq
where $0<x_1 < x_2 < ... < x_{N-1} < L$. (In a fermionic basis, the two 
sides of (\ref{bc1}) would differ by the phase $(-1)^{N-1} ~$). On the other
hand, if all the particles are distinguishable, we demand that the wave 
function should be periodic in each of the particles separately; namely,
\beq
\Psi (x_1 , ... , x_{i-1} , 0 , x_{i+1} , ... , x_N ) ~=~
\Psi (x_1 , ... , x_{i-1} , L , x_{i+1} , ... , x_N )
\label{bc2}
\eeq
whenever $0 < x_{i+1} < ... < x_N < x_1 < ... < x_{i-1} < L$. We can similarly
work out the appropriate boundary conditions if some of the particles are
distinguishable and some are not. In every case, we see that it is sufficient
to specify the wave function in the domain $0 \le x_1 < x_2 < ... < x_N 
\le L$; we will work in this particular domain from now on. The wave 
function in all other domains can be obtained by using the appropriate 
boundary conditions. It is important to distinguish between the concepts of 
sectors and domains. For instance, in the above example with four particles, 
the sector $(1122)$ has four domains while the sector $(1212)$ has two domains.
Under time evolution, particles can move from one domain to another. For any 
$N$ and any sector, it can be shown that $N$ is exactly divisible by the 
number of domains $D$. Further, a sector can be specified by repeating a 
pattern of $D$ integers $N/D$ times; the $D$ integers are chosen from the set 
$\{ 1,2,...,A \}$. For a single species CS model, there is only one sector and
one domain.

For $N=2$, there is only one sector; the number of domains is two and one if
the particles are distinguishable and indistinguishable respectively. The 
energy spectrum can be found exactly 
by separating the wave function into functions of the \com and 
relative coordinates; however we will see that there is an interesting 
selection rule if the particles are indistinguishable. The wave function 
$\Psi$ is labelled by two integers $n_1$, the \com quantum number 
which runs over all integers, and $n_2$, the relative coordinate number which 
must be zero or positive. In terms of the coordinates $X=(m_1 x_1 + m_2 x_2)/
(m_1 + m_2)$ and $x=x_2 - x_1$, we have
\beq
\Psi_{n_1 n_2} ~=~ \exp ~(~i 2 \pi n_1 {X \over L}~) ~~\vert ~ \sin ~({{\pi x}
\over L} ) ~\vert^{g_{12}} ~~Q_{n_2} (~\cos ~({{\pi x} \over L})~) ~,
\label{8p2}
\eeq
where $Q_{n_2}$ is a polynomial of degree $n_2$; it is even or odd depending 
on $n_2$. The energy is
\beq
E_{n_1 n_2} ~=~ {{\pi^2 \hbar^2} \over {L^2}} ~\Bigl[ ~{{2 n_1^2} \over 
{m_1 + m_2}} ~ +~ (n_2 ~+~ g_{12} )^2 ~{{m_1 +m_2} \over {2m_1 m_2}} ~\Bigr]~.
\label{8p3}
\eeq
For indistinguishable particles, we set $m_1 = m_2$ and $g_{12} = g_{11}$ in
Eqs. (\ref{8p2}-\ref{8p3}); in addition, we must restrict 
$n_1 + n_2$ to be even (or odd, depending on the choice of basis). To see this, 
we note that $\Psi (0,y) = (-1)^{n_1 + n_2} \Psi (y,L)$. In a bosonic basis, 
$n_1 +n_2$ must be even. In a fermionic basis, $n_1 +n_2$ must be odd. Note 
that the statement that the two particles are bosons (or fermions) if 
$g_{11} =0$ (or $1$) is independent of the choice of basis. 

For $N \ge 3$, the \gs of Eq. (\ref{ham1}) cannot be found exactly in 
general. It is also not clear that the \gs energy will be the same in all 
sectors. However the situation simplifies dramatically if we impose the 
relations (\ref{gm2}). The exact \gs wave function is then given by
\beq
\Psi_o ~=~ \prod_{i<j} ~\vert ~\sin ~[~{\pi \over L} ~(x_i - x_j) ~] ~
\vert^{g_{ij}} ~.
\label{9}
\eeq
This is the \gs since it has no nodes for $x_i \ne x_j$. The energy 
is exactly the same as in (\ref{3p5}). It is clear that the energy is the 
same in all sectors. 

If we write an eigenstate of Eq. (\ref{ham1}) in the form $\Psi = \Phi \Psi_o$,
then $\Phi$ must satisfy the eigenvalue equation
\beq
- ~[~\sum_i ~{{\hbar^2} \over {2m_i}} ~\partial_i^2 ~+~ {{\pi \hbar^2 \alpha} 
\over L} ~ \sum_{i<j} ~\cot [{{\pi} \over L} (x_i - x_j)] ~(m_j \partial_i ~-~ 
m_i \partial_j ) ~] ~\Phi ~=~ (E - E_o) ~\Phi ~,
\label{ham3}
\eeq
where $\partial_i = \partial / \partial x_i$. A class of exactly solvable
eigenstates is given by the excitations of the centre of mass; these are 
discussed in detail in the next paragraph. We have not been able to find any 
other excited states exactly; however we will discuss some approximate 
low-energy eigenstates in Sec. 6.

We will now study the \com excitations assuming that the realative coordinates
problem is in its ground state (\ref{9}). In our preferred domain $0 \le x_1 < 
x_2 < ... < x_N \le L$, the \com excitations have the wave function
\beq
\Phi ~=~ \exp ~(i ~{{2\pi n} \over {LM}} ~\sum_i m_i x_i ) ~, 
\label{com1}
\eeq
where $M= \sum_i m_i$ is the total mass. We now have to find the allowed
values of $n$. To do that, we have to consider all the domains and boundary
conditions together. For a given sector, let us label the different domains
by an integer $d=1,2,...,D$, where $d=1$ denotes our preferred domain. The 
functional form of the wave function in any domain $d$ is given by
\beq
\Phi_d ~=~ \Phi_1 ~\exp ~(i \theta_d) ~,
\label{com2}
\eeq
where $\Phi_1$ is defined in (\ref{com1}) and $\theta_1 =0$. The boundary
conditions now imply that if we go from the domain $1$ to a domain $2$ by 
moving only one particle $i$ across the point $x=0$ from $L^-$ to $0^+$, then
\beq
\theta_2 ~-~ \theta_1 ~=~ 2 \pi n ~{{m_i} \over M}
\eeq
(There is no extra phase arising from the relative coordinates wave function
(\ref{9}) since it remains unchanged if a coordinate $x_i$ changes by $L$). 
We can then go to a domain $3$ by moving another particle across $x=0$; if we 
repeat this process $D$ times, we come back to domain $1$ with the wave 
function $\Phi_1$ times a phase. We can show that the total phase accumulated 
is $2 \pi n D /N$.  Hence $n D /N$ must be an integer. We therefore have the 
important result that the spectrum of \com excitations
\beq
\Delta E_n ~=~ \Bigl({{2 \pi \hbar} \over L} \Bigr)^2 ~{{n^2} \over {2M}}
\label{com3}
\eeq
depends on the sector through the number of domains $D$. We would like to 
emphasize that this discussion of the \com excitations holds for a
{\it general} mulitspecies CS model in which the interaction parameters may
not satisfy (\ref{gm2}); this is because the dynamics of the \com
does not depend on the $g_{ij}$'s.

The \gs of the multispecies CS model is exactly solvable if we retain 
(\ref{gm2}) but consider the problem on a line in the presence of a simple 
harmonic potential, with the {\it same} harmonic frequency $\omega$ for all 
the particles \cite{KRI}. The Hamiltonian is 
\beq
H ~=~ \sum_i ~{{p_i^2} \over {2m_i}} ~+~ \sum_{i<j} ~g_{ij} (g_{ij} -1)~ 
{{m_i + m_j} \over {2m_i m_j}} ~{{\hbar^2} \over {(x_i - x_j)^2}} ~+~ 
{{\omega^2} \over 2} ~\sum_i ~m_i x_i^2 ~.
\label{ham2}
\eeq
The number of different sectors now is $N !\ / (\prod_a N_a !\ )$. In any 
sector, the \gs wave function is 
\beq
\Psi_o ~=~ \prod_{i<j} ~\vert x_i - x_j \vert^{g_{ij}} ~\exp ~[ - ~{\omega
\over {2\hbar}} ~\sum_i ~m_i x_i^2 ~] ~,
\label{12}
\eeq
and the energy is
\beq
E_o ~=~ {{\hbar \omega \alpha} \over 2} ~\Bigl( ~\sum_a ~m_a N_a ~
\Bigr)^2 ~+~ {\hbar \omega \over 2} ~\sum_a ~(~1 ~-~ \alpha ~m_a^2 ~)~ 
N_a ~.
\label{13}
\eeq

We can find an infinite number of excited states (but not all the excited
states) as follows. First of all, the Hamiltonian separates into a \com 
problem with coordinate $X = \sum_i ~m_i x_i ~/M$, and a relative 
coordinate problem. The latter further separates into one `radial' coordinate 
$r$ where
\beq
M r^2 ~=~ {1 \over M} \sum_{i<j} ~ m_i m_j ~(x_i - x_j )^2 ~,
\label{14}
\eeq
(this denotes the moment of inertia about the centre of mass) and $N-2$
`angular' coordinates collectively denoted by $\Omega_i$. The term inside the 
exponential in Eq. (\ref{12}) is $\sum_i m_i x_i^2 = M(X^2 + r^2)$. The 
Hamiltonian can be written as
\begin{eqnarray}
H ~=~ &-& ~{{\hbar^2} \over {2M}} ~{{\partial^2} \over {\partial X^2}} ~+~ 
{1 \over 2}~ M \omega^2 X^2 \nonumber \\
&-& ~{{\hbar^2} \over {2M}} ~[ ~{{\partial^2} \over {\partial r^2}} ~+~ {{N-2} 
\over r} ~ {\partial \over {\partial r}} ~] ~+~ {1 \over 2} ~M \omega^2 r^2 ~
+~ {{\hbar^2} \over {2 M r^2}}~ {\cal L} ~, 
\label{15}
\end{eqnarray}
where ${\cal L}$ is a differential operator which only acts on functions of 
$\Omega_i$. The eigenvalues of the \com Hamiltonian are given by
$\hbar \omega (n_1 + 1/2)$; the corresponding wave functions take the form
\beq
\Psi_{n_1} (X )~=~ P_{n_1} ({\sqrt {{M \omega} \over {\hbar}}} X) ~\exp ~
(~- {{M \omega} \over {2\hbar}} X^2 ~) ~,
\label{16}
\eeq
where $P_{n_1}$ denotes a polynomial of degree $n_1$. The entire difficulty in
finding the spectrum of (\ref{15}) lies in finding eigenstates $Y(\Omega_i)$ of 
$\cal L$ which 
satisfy the boundary conditions at coincident points and the symmetry
between particles of the same species. For any eigenstate of $\cal L$ with 
eigenvalue $\lambda (\lambda + N -3)$, where $\lambda > 0$, we get an 
infinite tower of eigenvalues of the relative coordinate Hamiltonian of the
form $\hbar \omega (2n_2 + \lambda + (N-1)/2)$. The $r$-dependence of the
corresponding wave function takes the form
\beq
\Psi_{n_2} (r) ~=~ r^\lambda ~Q_{n_2} ({{M \omega} \over {\hbar}} r^2) ~\exp ~
(~- {{M \omega} \over {2 \hbar}} r^2 ~)~,
\label{17}
\eeq
where $Q_{n_2}$ is a polynomial of degree $n_2$; it has $n_2$ nodes. 
The excitations corresponding to $Q_{n_2}$ are called breathing modes. In 
particular, a set of {\it exact} eigenstates of (\ref{15}) is given by 
\beq
\Psi_{n_1 n_2} ~=~P_{n_1} ({\sqrt {{M \omega} \over {\hbar}}} X) ~Q_{n_2} ({{M 
\omega} \over {\hbar}} r^2) ~ \Psi_o ~,
\label{18}
\eeq
and the energy is $E=E_o + \hbar \omega ( n_1 + 2 n_2 )$.

The Hamiltonians (\ref{ham1}) and (\ref{ham2}) can be written as 
\beq
H ~=~ \sum_i ~Q_i^\dagger Q_i ~+~ E_o ~, 
\label{19}
\eeq
where
\beq
Q_i ~=~ {1 \over {\sqrt {2m_i}}} ~[~p_i ~+~ {{i \pi \hbar} \over L} ~\sum_{j 
\ne i}~ g_{ij} ~\cot {\pi \over L} ~(x_i - x_j) ~]
\label{20}
\eeq
in (\ref{ham1}), and
\beq
Q_i ~=~ {1 \over {\sqrt {2m_i}}} ~[~p_i ~+~ i \hbar ~\sum_{j \ne i}~ {{g_{ij}} 
\over {x_i - x_j}} ~-~ i m_i \omega x_i ~]
\label{21}
\eeq
in (\ref{ham2}). The states (\ref{9}) and (\ref{12}) are annihilated by all the 
$Q_i$ in (\ref{20}) and (\ref{21}) respectively.

Incidentally, one can check that some of the assumptions made in Sec. 2 are
indeed valid for the CS model discussed here. Firstly, for two particles on a 
line, with the Hamiltonian being given by (\ref{ham2}) but without the 
simple harmonic confinement, we can solve for the scattering problem after 
going to relative coordinates. We find that the scattering phase shift is 
given by $\pi g_{ij} \theta (v_i - v_j)$. Secondly, if all the parameters 
$g_{ij}$ are integers, we can expand the \gs wave function (\ref{9}); 
it consists of a superposition of plane waves with different values of the 
particle momenta. We then discover that all the solutions of Eq. (\ref{dist2})
appear in the superposition. There are also terms in the superposition which 
do not correspond to solutions of (\ref{dist2}); however this is true even for 
the single species CS model.

\vskip .7 true cm
\leftline{\bf 4. Thomas-Fermi Calculation of the One-Particle Densities}
\vskip .3 true cm

In this section, we will use a generalized Thomas-Fermi (TF) method \cite{SEN} 
to obtain the one-particle densities $\rho_a (x)$ in the \gs of the 
CS model in an external harmonic potential. The advantage of using the TF
method is that one does not need to know the exact wave function; for the
single species CS model, the semicircle law comes out quite easily. For the
multispecies model, we will discover that the particle densities depend on the 
sector. 

In the absence of an external potential, Eq. (\ref{4}) says that
\beq
{{p_{Fa}} \over {m_a}} ~=~ \pi \hbar \alpha ~\sum_b ~m_b ~\rho_b
\label{22}
\eeq
is independent of both $a$ and $x$ in all sectors. In the presence of a 
potential, the TF method uses Eq. (\ref{22}) in a local sense, i.e., $p_{Fa}$ 
and $\rho_a$ become functions of $x$. The most energetic particle of species 
$a$ at position $x$ must satisfy
\beq
{{p_{Fa}^2} \over {2m_a}} ~+~ {1 \over 2} m_a \omega^2 x^2 ~=~ \mu_a ~,
\label{23}
\eeq
where $\mu_a$ is the global Fermi energy for $a$ particles. ($\mu_a$ must be
independent of $x$, otherwise one could move an $a$ particle from a position 
with a higher value of $\mu_a$ to a position with lower $\mu_a$. This would 
give us a different sector with a lower energy which is not possible). We thus
see that
\beq
\pi \hbar \alpha ~\sum_b ~m_b ~\rho_b (x) ~=~ \Bigl( ~{{2 \mu_a} \over 
{m_a}}~ -~ \omega^2 x^2 ~\Bigr)^{1/2}
\label{24}
\eeq
must be independent of $a$. Clearly, the particle densities must vanish outside
an interval $[-x_o , x_o]$, where
\beq
x_o^2 ~=~ {{2 \mu_a} \over {m_a \omega^2}} ~=~ {{2 \hbar \alpha} \over
\omega} ~\sum_b ~m_b ~N_b ~.
\label{25}
\eeq

Eq. (\ref{24}) only gives us a weighted sum of the particle densities, not the 
individual values of $\rho_a$. This is because the individual densities on the 
left hand side of (\ref{24}) depend on the ordering of the particles, although 
the right hand side is the same for all orderings. 

If we are given a particular ordering, we can compute the densities as follows.
Introduce a parameter $\tau$, where $0 \le \tau \le 1$, and functions $f_a 
(\tau)$ such that the first $\tau N$ particles from the left contain $f_a 
(\tau) N_a$ particles of species $a$. ($N$ is the total number of particles). 
We assume that $f_a (\tau)$ varies smoothly with $\tau$. Clearly, $f_a (0) = 
0$, $f_a (1) = 1$, and $f_a^\prime (\tau) \ge 0$. We emphasize that the 
functions $f_a$ are fixed once the ordering of the particles (i.e., a sector) 
is specified. Now we want to express $\tau$ as a function of $x$. At $x=-x_o$ 
and $x_o$, $\tau = 0$ and $1$ respectively. In the interval $x$ to $x+dx$ or, 
equivalently, $\tau$ to $\tau + d\tau$, the number of $a$ particles is equal to
\beq
\rho_a (x) ~dx ~=~ N_a f_a^\prime (\tau) ~d\tau ~.
\label{26}
\eeq
From Eqs. (\ref{24}-\ref{26}), we deduce that
\beq
\omega ~ \int_{-x_o}^x ~dy ~(x_o^2 ~-~ x^2)^{1/2} ~=~ \pi \hbar \alpha ~
\sum_a ~m_a ~N_a f_a (\tau) ~,
\label{27}
\eeq
so that $\tau$ is implicitly a function of $x$. The densities then follow from 
Eq. (\ref{26}), $\rho_a (x) = N_a f_a^\prime (\tau) d \tau / dx $.

It is interesting to observe that Eq. (\ref{1}) is valid semiclassically even
in the presence of a simple harmonic potential. We simply use Eq. (\ref{4p2}) 
with $h_a =p^2 /2m_a ~+~ m_a \omega^2 x^2 /2$, and let $\epsilon \rightarrow 
0$ after calculating $\Delta D_a (\epsilon)$.

We can use the TF method to calculate the one-particle densities even if the
confining potential is not simple harmonic and the \gs is not exactly solvable.
If $V_a (x)$ is the one-body potential felt by a particle of species $a$, we 
can find the condition under which the \gs energy will be the same in all 
sectors. As in Eq. (\ref{23}), we want
\beq
{{p_{Fa}^2} \over {2m_a}} ~+~ V_a (x) ~=~ \mu_a ~
\label{27p0}
\eeq
with $\mu_a$ independent of $x$. From (\ref{22}), this implies that $(\mu_a -
V_a (x))/m_a$ should be the same for all species. We conclude that $d(V_a (x)/
m_a)/dx$ (i.e., the classical acceleration) should be independent of $a$.

\vskip .7 true cm
\leftline{\bf 5. High Temperature}
\vskip .3 true cm

The entire analysis so far has been at zero temperature. We now study the CS 
model at finite temperature. Two things are particularly important to check.

\noindent
(a) Are thermodynamic properties like the equation of state the same in all 
sectors (as is true at zero temperature)? We will argue that the properties 
probably depend on the sector at high temperature. 

\noindent
(b) Does the model show mutual statistics, if not exactly, then at least 
approximately in some range of temperatures?

In this section, we consider the equation of state of the system at high 
temperature. The pressure $P$ can then be expanded in terms of the densities
$\rho_a =N_a /L$. We will only be interested in the expansion upto quadratic
order, i.e., in the second virial coefficients. The grand canonical partition
function $Z$ has an expansion of the form
\beq
Z ~=~ 1~+~ \sum_a ~\zeta_a Z_a ~+~ \sum_a ~\zeta_a^2 Z_{aa} ~+~ \sum_{a<b} ~
\zeta_a \zeta_b Z_{ab} ~+~ ...~,
\label{27p1}
\eeq
where $\zeta_a$ denotes the fugacity of species $a$, $Z_a$ is the partition 
function of a single particle of species $a$ on a circle of length $L$, and
$Z_{ab} =Z_{ba}$ is the partition function of two particles of species $a$ 
and $b$. Next, the densities satisfy
\beq
\rho_a ~=~ {{\zeta_a} \over L} ~{{\partial \log Z} \over {\partial \zeta_a}}~.
\label{27p2}
\eeq
We then obtain
\beq
\beta P ~=~ \sum_a ~\rho_a ~+~ \sum_a ~A_{aa} ~\rho_a^2 ~+~ \sum_{a<b} ~
A_{ab} ~\rho_a \rho_b ~+~ ... ~,
\label{27p3}
\eeq
where $\beta = 1/k_B T$, and the second virial coefficients are given by
\begin{eqnarray}
A_{aa} &=& \lim_{L \rightarrow \infty} ~L~ \Bigl(~ {1 \over 2} ~-~ {{Z_{aa}}
\over {Z_a^2}}~ \Bigr) ~, \nonumber \\
A_{ab} &=& \lim_{L \rightarrow \infty} ~L~ \Bigl(~ 1 ~-~ {{Z_{ab}}
\over {Z_a Z_b}}~ \Bigr)~.
\label{27p4}
\end{eqnarray}
In the limit $L \rightarrow \infty$, we can use the Poisson resummation
formula 
\beq
\sum_{n=-\infty}^{\infty} ~\exp ~(~ -\pi n^2 /L^2 ) ~=~ L~ 
\sum_{n=-\infty}^{\infty} ~\exp ~(~- \pi n^2 L^2 )
\label{27p5}
\eeq
to show that $Z_a = L/\lambda_a$ plus exponentially small terms; $\lambda_a = 
(2 \pi \hbar^2 \beta /m_a)^{1/2}~$ is called the thermal wavelength. Using Eq. 
(\ref{8p3}), we can calculate $Z_{aa}$ and $Z_{ab}$. We finally obtain
\begin{eqnarray}
A_{aa} &=& (g_{aa} -{1 \over 2})~ {{\lambda_a} \over {\sqrt 2}} ~, \nonumber \\
A_{ab} &=& (g_{ab} -{1 \over 2})~ {\sqrt {\lambda_a^2 ~+~ \lambda_b^2}} ~.
\label{27p6}
\end{eqnarray}
As mentioned earlier, there is only sector for the two-particle problem on a
circle, but there is an exponentially large number of sectors in the 
thermodynamic limit $N \rightarrow \infty$. The virial coefficients calculated 
above must therefore be some kind of an average over the different sectors, 
although we do not know at this point if all sectors contribute to the average 
with equal weight.

Now there is one sector in which we can calculate many properties exactly; 
this is the sector defined at the end of Sec. 2 where the different species 
are segregated, with $N_a$ particles of species $a$ occupying a segment of 
length $L_a$. We may compute the second virial coefficients in this sector as 
follows. For a single species, we know from the above equations that the 
pressure of species $a$ is given by 
\beq
\beta P_a ~=~ {\tilde {\rho_a}} ~+~ A_{aa} ~{\tilde \rho_a}^2 ~+~ ... ~,
\label{28}
\eeq
where ${\tilde {\rho_a}} = N_a / L_a$. At equilibrium, the pressures exerted 
by different species must be equal. This enables us to eliminate the $L_a$ in 
terms of the total length $L$ and the densities $\rho_a$. We thus get
\beq
\beta P ~=~ \sum_a  ~\rho_a ~+~ \sum_a ~A_{aa} ~\rho_a^2 ~+~ \sum_{a<b} ~(
A_{aa} ~+~ A_{bb} ) ~\rho_a \rho_b ~+~ ... ~.
\label{28p1}
\eeq
We see that the virial coefficients $A_{ab}$ in this sector do not agree with
those in Eq. (\ref{27p6}) since the masses are unequal. Thus the equation 
of state seems to depend on the sector. Since the second virial coefficients
should be related to the statistics parameters \cite{MUR1,NAY}, we conclude 
that the model cannot realize \ms in all sectors at high temperature.

We now observe that the particle density $\sum_a \rho_a$ is independent of both
the position $x$ and the sector at $T \rightarrow \infty$, while the mass 
density $\sum_a m_a \rho_a$ is independent of the position and sector at $T 
\rightarrow 0$ (cf. Sec. 2). This implies that the individual densities 
$\rho_a$ must vary with the temperature in at least some sectors. Such a 
variation may 
seem surprising in a model which is scale invariant, in the sense that the
kinetic terms and the two-body interactions transform in the same way under an
uniform scaling of all the coordinates. However the temperature-dependence 
might be understandable if the model is found to be non-integrable. We have 
therefore checked for integrability, both classically and quantum 
mechanically, using the methods given in Refs. \cite{POL,MOS}; the model
appears to be non-integrable. Hence the asymptotic Bethe ansatz is not expected
to be valid for the excited states, except for the \com excitations 
in which all the velocities are shifted by the same amount from their \gs 
values.

We recall that in the {\it single} species CS model, the \es is ideal. This 
happens because the scattering phase shift is independent of the momenta of 
the two particles {\it and} the model is integrable \cite{POL,MOS}. The second 
property implies that an $N$-particle scattering phase shift is given by a sum 
of two-particle phase shifts.

Although the CS model does not seem to exhibit \ms in all sectors at high 
temperature, it may do so in an approximate way at low temperature. In the 
next section, we study the low-energy excitations in some detail.

\vskip .7 true cm
\leftline{\bf 6. Low Temperature}
\vskip .3 true cm

A general formalism for obtaining the distribution functions for a system with 
\ms is given in Ref. \cite{FUK}. It is therefore interesting to investigate, 
(a) whether the low-energy excitations are described by one or more $c=1$ 
Gaussian conformal field theories \cite{FUK}, and (b) whether the equation of 
state of the CS model at low temperature agrees with that of a system with 
mutual statistics. We will provide a complete answer to (a), but we are unable 
to answer question (b) by analytical methods.

The elementary excitations in a Gaussian theory satisfy the dispersion relation
\beq
\Delta E ~=~ v_F ~\vert \Delta p \vert
\label{e1}
\eeq
where $\Delta E$ and $\Delta p$ respectively denote the energy and momentum 
with respect to the ground state, and $v_F$ is the Fermi velocity. From Eq.
(\ref{fv}), $v_F = \pi \hbar \alpha M /L$, where $M=\sum_i m_i$ is the total 
mass; $v_F$ remains finite in the thermodynamic limit $M, L \rightarrow 
\infty$. (Actually, Eq. (\ref{e1}) describes the excitations of a 
Gaussian theory based on an {\it uncompactified} scalar field. For a 
compactified scalar field, there are some additional terms in the energy which 
will be discussed in Eq. (\ref{ep2}) below).

We can now find a right-moving elementary excitation approximately by using 
the Sutherland ansatz for the wave function \cite{SUT}. Let us first define
\beq 
z_i ~=~ \exp (i {{2\pi} \over L} x_i ) ~.
\eeq
Following Eq. (\ref{ham3}), we write 
$\Psi_1 = \Phi_1 \Psi_o$ where
\beq
\Phi_1 ~=~ \sum_i ~m_i ~z_i ~.
\label{phi1}
\eeq
This has $\Delta p = 2 \pi \hbar /L$ since the momentum operator is $p = 
\sum_i p_i$. We can show that the expectation value
\beq 
\Delta E ~=~ {{\langle \Psi_1 \vert H - E_o \vert \Psi_1 \rangle} \over 
{\langle \Psi_1 \vert \Psi_1 \rangle}} ~=~ v_F ~\Delta p ~+~ O({{\Delta p} 
\over L}) ~.
\label{deltae}
\eeq
The corrections of order $\Delta p/L$ arise due to the approximate nature of 
the wave function, and they go to zero as $L, N \rightarrow \infty$; we then 
recover $\Delta E / \Delta p = v_F$. Similarly, a left-moving elementary 
excitation with $\Delta p = - 2 \pi \hbar /L$ is given by $\Psi_{-1} =
\Phi_{-1} \Psi_o$ where $\Phi_{-1} = \Phi_1^{\star}$. It is important to note 
that the leading term in $\Delta E /\Delta p$ is independent of the sector.

The first term in $\Delta E/\Delta p$ in Eq. (\ref{deltae}), 
which survives as $L \rightarrow \infty$, arises entirely from the {\it linear}
differential operators in Eq. (\ref{ham3}). On the other hand, the corrections 
of order $1/L$ arise from both the linear and the quadratic differential 
operators in (\ref{ham3}). It is very useful to know that the leading term
in $\Delta E$ arises from a linear operator; it implies that
the energy eigenvalues will add up when we multiply two eigenstates together.
Thus a family of composite excitations with $\Delta p = 2 \pi \hbar (n_1 - 
n_2) /L$ is given by
\beq
\Psi_{n_1 , n_2} ~=~ \Phi_1^{n_1} ~\Phi_{-1}^{n_2} ~\Psi_o ~,
\label{phi2}
\eeq
where $n_1 , n_2 \ge 0$; this state has $\Delta E = v_F 2 \pi \hbar (n_1 + 
n_2)/L$. We require that $n_1 , n_2 << N$ so that this is a {\it low-energy} 
excitation and the corrections of order $n_i /L$ vanish as $L \rightarrow 
\infty$.

Following Ref. \cite{SUT}, we can show that the general excitation 
is specified by a set of non-zero integers $(n_1 , n_2 , ... , n_k)$; they may
be positive or negative. (For this to describe a low-energy excitation, 
we again require that each of the magnitudes $\vert n_i \vert$ as well as $k$
should be much less than $N$). The energy and momentum of such a state are 
given by
\bea
\Delta E ~&=&~ v_F ~{{2 \pi \hbar} \over L} ~\sum_i ~\vert n_i \vert ~,
\nonumber \\
\Delta p ~&=&~ {{2 \pi \hbar} \over L} ~\sum_i ~n_i ~.
\label{ep1}
\eea
If all the integers $n_i$ are positive, the ansatz wave function $\Phi_{n_1,
n_2,...,n_k}$ contains terms like $(m_1 z_1)^{n_1} (m_2 z_2)^{n_2} ... 
(m_k z_k)^{n_k}$ plus permutations, plus other terms which are obtained by 
squeezing (see below). If an integer $n_i$ is negative, $(m_i z_i)^{n_i}$ is 
replaced by $(m_i z_i^{\star})^{-n_i}$. The wave function $\Phi_{n_1,n_2,...,
n_k}$ is obtained by multiplying together the ansatz wave functions of a number 
of `elementary' states; an elementary state is one in which
all the integers $n_i$ are equal to $1$, or all of them are equal to $-1$.

Let us first examine the wave function for an elementary 
state, say, $(1,1)$. This is given by $\Psi_{1,1} = \Phi_{1,1} \Psi_o$ where
\beq
\Phi_{1,1} ~=~ \sum_{i<j} ~f(x_i - x_j) ~m_i m_j z_i z_j ~.
\label{phi3}
\eeq
The factor $f(x_i - x_j)$ in (\ref{phi3}) requires some explanation. If the
particles $i$ and $j$ belong to the same species, we set $f=1$. But if they 
belong to {\it different} species, we require that $f$ should go to zero 
sufficiently fast if $i$ and $j$ approach each other, and should go to $1$ if 
their separation is much larger than the inverse density $L/N$. For instance, 
we could define the function as
\beq
f(x_i - x_j) ~=~ \exp \Bigl( - {{\vert g_{ii} - g_{jj} \vert} \over {N \sin 
[{{\pi} \over L} (x_i - x_j)]}} \Bigl)
\eeq
The reason for introducing such a function in (\ref{phi3}) is that if we 
always set $f=1$, the linear operator in Eq. (\ref{ham3}) would act on
(\ref{phi3}) to produce a function which diverges when $x_i \rightarrow x_j$;
the expectation value $\langle \Psi \vert H \vert \Psi \rangle$ would then
be ill-defined if $g_{ij}$ is not large enough. (We would like to emphasize 
that the function $f$ is only required because we are working with 
{\it approximate} wave functions; such a short-distance cutoff 
is not present in the exact wave functions). Since $f$ only differs from
$1$ in certain microscopic regions of space, its presence can only change the 
expectation value of $\Delta E / \Delta p$ by an amount which vanishes as
$L \rightarrow \infty$. We can now verify that the energy and momentum of the
state (\ref{phi3}) is given by (\ref{ep1}). 

We can write down the wave functions of other elementary states like $(1,1,1)$ 
in a similar way. The wave functions of left-moving states such as $(-1,-1)$ 
are obtained by complex cojugating $\Phi_{1,1}$. Finally, composite wave 
functions such as $\Phi_{2,1,-1}$ can be written as a product of $\Phi_{1,1}$, 
$\Phi_1$ and $\Phi_{-1}$. The general rule for writing $\Phi_{n_1,n_2,...,
n_k}$ as a product of elementary wave functions can be expressed inductively 
as follows. Suppose that the numerically largest positive (or negative) 
integer $n_i$ appears $l$ times in the set $(n_1 , n_2 , ... , n_k)$. For 
instance, let $n_1 = n_2 = ...  = n_l$ be the largest positive integers. Then
\beq
\Phi_{n_1,n_2,...,n_l,n_{l+1},...,n_k} ~=~ \Phi_{n_1 -1,n_2 -1,...,n_l -1, 
n_{l+1},...,n_k} ~\Phi_{1,1,...,1} ~, 
\label{phin}
\eeq
where the last $\Phi$ has $l$ $1$'s.
This factorization can be repeated till we only get elementary wave functions 
on the right hand side of Eq. (\ref{phin}). If 
$n_1 = ... = n_l$ is negative, we replace $n_i - 1$ by $n_i + 1$, for $1 
\le i \le l$, and $\Phi_{1,1,...,1}$ by $\Phi_{-1,-1,...,-1}$ on the right 
hand side of (\ref{phin}). An important point to note in all this is that a 
factor of $m_i$ comes along with each $z_i$ or $z_i^{\star}$. 

Finally, we observe that $\Phi_{2,1,-1}$ contains within it other wave 
functions such as $\Phi_{1,1,1,-1}$, $\Phi_{1,1}$ and $\Phi_2$, because the 
sets $(1,1,1,-1)$, $(1,1)$ and $(2)$ can be obtained form the set $(2,1,-1)$ 
by a process called squeezing \cite{SUT}. We say that a set $\{n_i^{\prime} \}$
is obtained from another set $\{ n_i \}$ by squeezing if, either 

\noindent (a) an integer $n_i$ in the second set is equal to the sum of 
two integers $n_j^{\prime}$ and $n_k^{\prime}$ in the first set, where 
$n_i, n_j^{\prime}$ and $n_k^{\prime}$ all have the same sign, or 

\noindent (b) there are two integers $n_i , n_l$ in the second set, and {\it
upto} two integers $n_j^{\prime}$ and $n_k^{\prime}$ in the first set, which 
satisfy $n_i + n_l = n_j^{\prime} + n_k^{\prime}$ and $n_i > n_j^{\prime} , 
n_k^{\prime} > n_l$. (One or both of the integers $n_j^{\prime}$ and 
$n_k^{\prime}$ may be zero, in which case it is omitted from the set 
$\{ n_i^{\prime} \}$).

Following (\ref{ep1}), we can think of the low-energy excitations in terms of a 
system of `bosons' whose allowed energy levels are given by 
\beq
e_n ~=~ v_F ~{{2\pi \hbar} \over L}~ \vert n \vert ~,
\eeq
where $n$ can be any non-zero integer. These bosons have zero chemical 
potential. The partition function is therefore 
\beq
Z ~=~ \exp (-\beta E_o) ~\Bigl[ ~\prod_{n=1}^{\infty} ~(~1 - \exp (-\beta 
e_n)~) ~ \Bigl]^{-2} ~.
\eeq
We can now calculate the leading terms in the specific heat and pressure at 
low temperature. The mean energy is given by $E = - (\partial \ln Z /
\partial \beta)_L$, the specific heat is $C_V = (\partial E / \partial T)_L$, 
and the pressure is $P=k_B T (\partial \ln Z / \partial L)_T$. The leading 
terms in the specific heat and pressure are therefore given by
\bea 
{{C_V} \over L} ~&=&~ {{\pi k_B^2 T} \over {3 \hbar v_F}} ~, \nonumber \\
P ~&=&~ {{v_F^3} \over {3 \pi \hbar \alpha}} ~+~ {{\pi k_B^2 T^2} \over {3
\hbar v_F}} ~.
\label{cvp}
\eea
These leading terms are independent of the sector.
However we are unable to obtain the next term of order $T^2$ in $C_V$ or
order $T^3$ in $P$, since that requires a knowledge of the $1/L$ terms in the 
energy; we have ignored those terms because they do not seem to be 
analytically computable. The importance of the next term in $C_V$ or $P$ is 
that it contains information about the statistics parameters \cite{NAY}. 

Thus the low-energy excitations match those of a $c=1$ Gaussian conformal
field theory, rather than $c=A$ (as would be the case if all the off-diagonal
elements $g_{ab}$ were zero). This raises the following question. Are the low
temperature thermodynamic properties of the model the same as those of a {\it 
single} species CS model with some `effective' particle density $\rho$, mass 
$m$, and interaction parameter $g$? On comparing the specific heat and 
pressure in (\ref{cvp}) with those of a single species model, we see that the 
three effective parameters must satisfy the two relations
\bea
{{\pi \hbar g \rho} \over m} ~&=&~ v_F ~, \nonumber \\
{g \over {m^2}} ~&=&~ \alpha ~.
\eea
Clearly, we need a third low-energy property (for instance, the next term in 
$C_V$ or $P$) to fix the individual values of $\rho$, $m$ and, most importantly,
the \es parameter $g$. As mentioned earlier, we also have to determine whether 
or not these values depend on the sector. 

Before concluding this section, we note that the \com excitations discussed 
around Eq. (\ref{com3}) can have energies which are much {\it lower} than 
the ones discussed above if the number of domains $D$ in a given sector is
comparable to $N$. If the number of domains $D << N$ (as in a single species
CS model where $D=1$), the \com energies are of the same order as the ones
discussed above. In any case, the \com constitutes only a single
degree of freedom, so it has no effect on thermodynamic properties.

For the sake of completeness, we should mention that the low-energy
excitations of a single species CS model have some terms in the expressions 
for energy and pressure in addition to the ones given in (\ref{ep1}). These 
are characterized by two integers, $\Delta K$
equal to the number of particles transferred from the left Fermi momentum 
$-p_F$ to the right Fermi momentum $p_F$, and $\Delta N$ equal to the change 
in the number of particles. The contributions of these to the energy and
momentum take the form \cite{FUK}
\bea
\Delta E ~&=&~ \mu \Delta N ~+~ v_F ~{{2 \pi \hbar} \over L} ~\Bigl[ ~
{{(\Delta K)^2} \over g} ~+~ {g \over 4} (\Delta N)^2 ~\Bigr] ~, \nonumber \\
\Delta p ~&=&~ 2 \pi \hbar \rho ~\Delta K ~+~ {{2 \pi \hbar} \over L} ~\Delta K
\Delta N ~.
\label{ep2}
\eea
We observe that $\Delta K$ simply corresponds to a \com excitation
in which all the particle momenta are shifted by the same amount $2 \pi \hbar
\Delta K / L$; in the notation of Eq. (\ref{com3}), $\Delta K = n /N$. Eq. 
(\ref{ep2}) is essential in order to identify the radius of 
compactification of the scalar field in the Gaussian theory as $R = 1/ 
{\sqrt g}$. In the multispecies model, we can certainly calculate terms similar
to (\ref{ep2}) using Eqs. (\ref{com3}) and (\ref{3p5}); however we will not
exhibit them here since these terms have no bearing on the thermodynamics 
poperties which were calculated above with fixed particle numbers $N_a$.

\vskip .7 true cm
\leftline{\bf 7. A Multispecies Anyon Model in Two Dimensions}
\vskip .3 true cm

In this section, we would like to point out that relations similar to 
(\ref{gm2}) lead to some special properties for \ms in two dimensions. 
Consider a 
collection of $A$ species of anyons with \ms parameters $g_{ab}=g_{ba}$ and 
charges $q_a$ \cite{ISA1}. Let us first assume that all the $q_a$ 
have the same sign and all the $g_{ab}$ are positive. We now apply an uniform 
magnetic field in a direction normal to the plane. At zero temperature, all
the particles lie in the lowest Landau level (LLL). Let $B q_a > 0$. Then in 
the symmetric gauge $(A_x , A_y)=(-By/2,Bx/2)$, all the wave functions in the
LLL have the form 
\beq
\Psi ~=~ f(z_i) ~ \exp \Bigl[ - {B \over {4c \hbar}} ~\sum_i ~q_i {\vec r}_i^{~
2} ~ \Bigr] ~,
\label{30}
\eeq
where $z_i = x_i + i y_i$ and ${\vec r}_i^2 = z_i z_i^\star$; $c$ is the speed 
of light. (Note that the particle masses play no role in the LLL). When anyon 
$i$ is taken around a closed loop surrounding anyon $j$, the wave function 
should pick up a phase $\exp (i2 \pi g_{ij})$. The simplest wave function of 
this kind is given by 
\beq
f(z_i) ~=~ \prod_{i<j} ~(z_i ~-~ z_j)^{g_{ij}} ~.
\label{31}
\eeq
This is multivalued if any of the $g_{ij}$ is not equal to $0$ or $1$. Other 
wave functions in the LLL are obtained by multiplying (\ref{31}) by any 
polynomial in the $z_i$ which is symmetric between anyons of the same species. 
We will only consider the wave function (\ref{30}-\ref{31}) henceforth. 
Amongst all the LLL wave functions, this has the lowest angular momentum, i.e.,
\beq
L ~=~ \hbar \sum_i ~(~z_i {\partial \over {\partial z_i}} ~-~ z_i^\star 
{\partial \over {\partial z_i^\star}} ~)
\label{32}
\eeq
is equal to $\hbar \sum_{i<j} ~g_{ij}$. The particles with this wave function 
lie closest to the origin $z_i = 0$. 

The probability density is given by 
\beq
\vert \Psi \vert^2 ~=~ \prod_{i<j} ~\vert ~{\vec r}_i - {\vec r}_j ~\vert^{2
g_{ij}} ~\exp ~\Bigl[~ - {B \over {2c \hbar}} ~\sum_i ~q_i {\vec r}_i^{~2} ~
\Bigl]
\label{33}
\eeq
We now note that the expression (\ref{33}) may sometimes be valid even in a
situation in which the wave function is {\it not} given by (\ref{30}); in that 
case, $q_i$ in (\ref{33}) will denote the magnitude of the charge of species
$i$, rather than the charge itself. This is in fact true for the example given 
at the end of this section.

We will calculate the one-particle densities $\rho_a ({\vec r})$ using a 
plasma analogy \cite{LAU}. Since (\ref{33}) is circularly symmetric, $\rho_a$ 
can only be a function of the magnitude
$r$. Now $- \ln \vert \Psi \vert^2$ can be interpreted as the potential
energy of a system of point charges interacting with each other through
a repulsive Coulomb potential (which is logarithmic in two dimensions) and
with an uniform attractive background charge density equal to $-B/\pi c 
\hbar$. A particle of species $a$ which is at a distance $r$ from the origin 
will be at equilibrium if
\beq
\sum_b ~2g_{ab} ~{{Q_b (r)} \over r} ~=~ {B \over {c \hbar}} ~q_a r ~,
\label{34}
\eeq
where 
\beq
Q_b (r) ~=~ 2 \pi ~\int_0^r ~r^\prime dr^\prime ~\rho_b (r^\prime)
\label{35}
\eeq
is the total charge of species $b$ inside the circle of radius $r$. Eqs.
(\ref{34}-\ref{35}) imply that 
\beq
\pi ~\sum_b ~g_{ab} ~\rho_b (r) ~=~ {B \over {2c \hbar}} ~q_a ~.
\label{36}
\eeq
This is a system of $A$ equations if all the densities $\rho_a$ are non-zero
at $r$; if some of them are zero, we have to drop the corresponding equations
from (\ref{36}).

We now demand that the system should satisfy the following requirements. 
Firstly, since the right hand side of (\ref{36}) does not depend on $r$, we 
would like the particle distribution to be homogeneous, that is, $\rho_a$ 
should be independent of $r$. Secondly, if the numbers of particles $N_a$ are 
large but finite, we demand that there should be a single radius $R >> {\sqrt 
{c \hbar /Bq_a}}$ such that {\it all} the densities $\rho_a$ are non-zero 
inside $R$ and zero outside $R$. These two requirements rule out the 
possibility of phase separation, namely, a situation in which some, but not 
all, of the densities vanish in some region. We then have
\beq
\sum_b ~g_{ab} ~N_b ~=~ {{B R^2} \over {2c \hbar}} ~q_a ~.
\label{37}
\eeq
Finally, we demand that the numbers $N_a$ be independent variables; for
instance, $N_1$ should not get fixed if $N_2 , N_3, ... , N_A$ are specified.
We can show that these three requirements will be satisfied if and only if all 
the equations in (\ref{37}) are linearly dependent on just one of them, i.e., if
\beq
g_{ab} = \alpha ~q_a q_b 
\label{38}
\eeq
for both $a=b$ and $a\ne b$.

It is interesting to note that the relations in (\ref{38}) arise quite 
naturally in the fractional quantum Hall effect. For instance, consider the
situation when the filling fraction is close to but
slightly less than $1/m$, where $m$ is an odd integer like $3$ or $5$. In the 
first hierarchy, the ground state contains two species of particles with 
opposite charges, namely, the electron (species $1$) and the quasihole 
(species $2$). Laughlin's wave function then tells us that $g_{11} =m$, 
$g_{12} =1$, $g_{22}=1/m$, and the ratio of the magnitudes of the charges is
$q_2/q_1 = 1/m$ \cite{LAU}; hence (\ref{38}) is satisfied. Ref. \cite{ISA1} 
points out the interesting fact a fractional quantum Hall system can be 
described by a {\it single} Chern-Simons gauge field only if Eqs. (\ref{38}) 
are satisfied.

\vskip .7 true cm
\leftline{\bf 8. Discussion}
\vskip .3 true cm

To conclude, we have argued that certain relations between the interaction
parameters and masses are {\it necessary} for the \gs of a 
one-dimensional scale invariant multispecies model to satisfy the asymptotic 
Bethe ansatz. We have shown that a CS model with those relations has an 
exactly solvable ground state, and that the low-energy excitations are the 
same as those of a Gaussian conformal field theory. However we have not proved 
that the same relations are {\it sufficient} to give rise to either mutual 
statistics, or to a single species exclusion statistics. To address that 
question, we need to understand the spectrum of low-energy excitations more 
accurately.

\vskip .7 true cm
\leftline{\bf Acknowledgments}
\vskip .3 true cm

I thank R. K. Bhaduri, P. Durganandini, S. Mashkevich and M. V. N. Murthy 
for stimulating comments.


\newpage

\begin{thebibliography}{99}

\bibitem{CAL}F. Calogero, J. Math. Phys. 10, 2191, 2197 (1969).

\bibitem{SUT}B. Sutherland, J. Math. Phys. 12, 246, 251 (1971); Phys. Rev. A 
4, 2019 (1971); Phys. Rev. A 5, 1372 (1972).

\bibitem{POL}A. P. Polychronakos, Phys. Rev. Lett. 69, 703 (1992); B. 
Sutherland and B. S. Shastry, Phys. Rev. Lett. 71, 5 (1993). 

\bibitem{MOS}J. Moser, Adv. Math. 16, 197 (1975); F. Calogero, Lett. Nuovo 
Cimento 13, 411 (1975).

\bibitem{MEH}M. L. Mehta, Random Matrices (Academic Press, New York, 1991);
B. D. Simons, P. A. Lee and B. L. Altshuler, Phys. Rev. Lett. 72, 64 (1994).

\bibitem{HAL1}F. D. M. Haldane, Phys. Rev. Lett. 67, 937 (1991). 

\bibitem{DAS}A. Dasnieres de Veigy and S. Ouvry, Phys. Rev. Lett. 72, 600 
(1994); Mod. Phys. Lett. A 10, 1 (1995).

\bibitem{ISA1}S. B. Isakov, S. Mashkevich and S. Ouvry, Nucl. Phys. B 448, 
457 (1995).

\bibitem{MUR1}M. V. N. Murthy and R. Shankar, Phys. Rev. Lett. 72, 3629 (1994).

\bibitem{WU}Y.-S. Wu, Phys. Rev. Lett. 73, 922 (1994).

\bibitem{BER}D. Bernard and Y.-S. Wu, Proceedings of the 6th Nankai workshop 
on New developments of integrable systems and long-ranged interaction 
models, eds. M. L. Ge and Y.-S. Wu (World Scientific, Singapore, 1995).

\bibitem{ISA2}S. B. Isakov, Phys. Rev. Lett. 73, 2150 (1994); Int. J. Mod. 
Phys. A9, 2563 (1994); Mod. Phys. Lett. B 8, 319 (1994).

\bibitem{MUR2}M. V. N. Murthy and R. Shankar, Phys. Rev. Lett. 73, 3331 (1994).

\bibitem{FUK}T. Fukui and N. Kawakami, Phys. Rev. B 51, 5239 (1995); J. Phys. 
A28, 6027 (1995); N. Kawakami and S.-K. Yang, Phys. Rev. Lett. 67, 2493 (1991);
A. G. Izergin, V. E. Korepin and N. Yu. Reshetikhin, J. Phys. A22, 2615 (1989).

\bibitem{NAY}C. Nayak and F. Wilczek, Phys. Rev. Lett. 73, 2740 (1994); S. B. 
Isakov, D. P. Arovas, J. Myrheim and A. P. Polychronakos, Phys. Lett. A 212,
299 (1996).

\bibitem{HAL2}F. D. M. Haldane, Phys. Rev. Lett. 60, 635 (1988); B. S. 
Shastry, ibid. 60, 639 (1988).

\bibitem{FUR}C. Furtlehner and S. Ouvry, Mod. Phys. Lett. B 9, 503 (1995).

\bibitem{KRI}V. Ya. Krivnov and A. A. Ovchinnikov, Theor. Math. Phys. 50, 
100 (1982); P. J. Forrester, J. Phys. A 25, L607, 5447 (1992).

\bibitem{YAN}C. N. Yang and C. P. Yang, J. Math. Phys. 10, 1115 (1969).

\bibitem{SEN}D. Sen and R. K. Bhaduri, Phys. Rev. Lett. 74, 3912 (1995).

\bibitem{LAU}R. B. Laughlin, in The Quantum Hall Effect, eds. R. E. Prange 
and S. M. Girvin (Springer-Verlag, Berlin, 1986).

\end{thebibliography}
\end{document}